\PassOptionsToPackage{table,dvipsnames}{xcolor}
\documentclass[sigconf]{acmart}

\AtBeginDocument{%
  }
    
\copyrightyear{2026}
\acmYear{2026}
\setcopyright{cc}
\setcctype{by-nc-nd}
\acmConference[CCS '26]{Proceedings of the 2026 ACM SIGSAC Conference on Computer and Communications Security}{November 15--19, 2026}{The Hague, Netherlands}
\acmBooktitle{Proceedings of the 2026 ACM SIGSAC Conference on Computer and Communications Security (CCS '26), November 15--19, 2026, The Hague, Netherlands}
\acmISBN{979-8-4007-2871-6/2026/11}
\acmDOI{10.1145/3830454.3832676}

\usepackage{scalerel}
\usepackage[normalem]{ulem}
\usepackage{multirow}
\usepackage{adjustbox}
\usepackage{mathtools}
\usepackage{listings}
\usepackage{balance}
\usepackage{stmaryrd}

\usepackage{thmtools}
\usepackage{thm-restate}

\usepackage[%
  capitalize,
  noabbrev,
  nameinlink
]{cleveref}

\crefname{equation}{}{}
\Crefname{equation}{}{}

\input{macros-meta.tex}
\newcommand{\obsmodel}{\mathcal{M}}
\newcommand{\obs}{\mathcal{O}}

\newcommand{\ctobsmodel}{\mathcal{M}_{\mathit{ct}}}
\newcommand{\specobsmodel}{\mathcal{M}_{\mathit{spec}}}

\newcommand{\birobs}{\fstlangcol{{\obs^\biridx}}}

\newcommand{\cnd}[1]{\fstlangcol{Cnd}({#1})}
\newcommand{\ld}[1]{\fstlangcol{Ld}({#1})}
\newcommand{\st}[1]{\fstlangcol{St}({#1})}
\newcommand{\opc}[1]{\fstlangcol{Pc}({#1})}
\newcommand{\oempty}{\fstlangcol{\varepsilon}}

\newcommand{\mstates}{S}
\newcommand{\mstate}{s}

\newcommand{\exectrace}[1]{\tau^{#1}}
\newcommand{\exectraces}{T}

\newcommand{\functionsymbol}[1]{\mathlabel{#1}}

\input{macros-bir.tex}
\newcommand{\CryptoBap}{{\sc CryptoBap}}

\newcommand{\tree}{\fstlangcol{T}}
\newcommand{\node}{\fstlangcol{node}}
\newcommand{\branchingnode}{\fstlangcol{Branch}}
\newcommand{\leafnode}{\fstlangcol{Leaf}}
\newcommand{\nodecond}{\fstlangcol{\phi}}
\newcommand{\nodeevent}{\fstlangcol{ev}}

\newlength\shlength
\newcommand\xshlongvec[2][0]{\setlength\shlength{#1pt}%
	\stackengine{-5.6pt}{$#2$}{\smash{$\kern\shlength%
			\stackengine{7.55pt}{$\mathchar"017E$}%
			{\rule{\widthof{$#2$}}{.57pt}\kern.4pt}{O}{r}{F}{F}{L}\kern-\shlength$}}%
	{O}{c}{F}{T}{S}}

\input{macros-parallel.tex}

\newcommand{\Funcs}{\mathcal{F}}

\newcommand{\Term}{\mathcal{T}}

\newcommand{\Terms}{\ensuremath{\mathcal{T}}\xspace}

\newcommand{\pout}{\sndlangcol{\mathsf{out}}\xspace}
\newcommand{\pin}{\sndlangcol{\mathsf{in}}\xspace}
\newcommand{\pif}{\sndlangcol{\mathsf{if}}\xspace}
\newcommand{\pthen}{\sndlangcol{\mathsf{then}}\xspace}
\newcommand{\pelse}{\sndlangcol{\mathsf{else}}\xspace}
\newcommand{\plet}{\sndlangcol{\mathsf{let}}\xspace}
\newcommand{\pnew}{\sndlangcol{\mathsf{new}}\xspace}

\newcommand{\pevent}{\sndlangcol{\mathsf{event}}\xspace}

\newcommand{\traces}{\ensuremath{\mathit{traces}}\xspace}

\newcommand{\vars}{\ensuremath{\mathit{vars}}}

\newcommand{\theactualrule}[1]{\text{Please redefine the command
theactualrule.}}
\newcommand{\underscorethingy}[1]{\text{Please redefine the command
underscorethingy.}}

\begin{document}

\title{Automated Side-Channel Analysis of Cryptographic Protocol Implementations}%
\author{Faezeh Nasrabadi}
\affiliation{%
 \institution{CISPA Helmholtz Center for Information Security}
 \city{Saarbrücken}
 \country{Germany}
}\email{faezeh.nasrabadi@cispa.de}

\author{Robert Künnemann}
\affiliation{%
 \institution{CISPA Helmholtz Center for Information Security}
 \city{Saarbrücken}
 \country{Germany}
 \orcid{0000-0003-0822-9283}
}\email{robert.kuennemann@cispa.de}

\author{Hamed Nemati}
\affiliation{%
  \institution{KTH Royal Institute of Technology}
 \city{Stockholm}
 \country{Sweden}
 \orcid{0000-0001-9251-3679}
}\email{hnnemati@kth.se}

	\begin{abstract}
		Formal verification of cryptographic protocols typically relies on symbolic models that abstract away compiled code and microarchitectural side channels, leaving a gap between verified specifications and deployed executables.
We present a toolchain that extracts protocol-relevant models from real binaries and analyzes them under explicit leakage contracts for constant-time and Spectre-PHT-style speculative observations.
Starting from a selected binary region, we lift machine code to an intermediate representation, instrument it with leakage contracts, symbolically execute it to obtain event/observation traces, and translate these traces into \Sapic for analysis with \Tamarin, \ProVerif, and \DeepSec.

As case studies, we extract models of WhatsApp Desktop's session-management and double-ratchet components from its binary and analyze forward secrecy and post-compromise security under a state-cloning compromise.
For side-channel analysis, we study the Basic Access Control (BAC) protocol used in e-passports and WhatsApp's session establishment.
Under our observation models, we identify an instruction-cache side channel in WhatsApp Desktop enabling social-graph inference, and we reproduce known unlinkability issues in BAC under microarchitectural observations.

	\end{abstract}

	\maketitle
	\section{Introduction}
\label{sec:intro}

Cryptographic protocols form the backbone of modern digital security, protecting sensitive data across online interactions. Yet their design and implementation remain prone to subtle errors
both
at the specification level (e.g., the POODLE attack on SSL version 3.0, which exploited SSL's use of CBC-mode encryption with predictable padding~\cite{POODLE}) and at the implementation level (e.g., Heartbleed, \texttt{CVE-2014-0160}).
Formal methods can provide a systematic framework for finding attacks
and even ensuring the absence of (defined classes of) them.
But to find attacks on both levels, we need models that capture the
behavior of the compiled machine code, which is what ultimately runs on hardware.
Overcoming this \emph{gap} promises to increase
the trustworthiness of analyses, but also seems necessary
considering the rising complexity of modern protocols, which often
involve various sub-protocols like session management and key
exchange.
It also offers an opportunity to further investigate and identify attacks resulting from hardware interaction.
In particular, \emph{side-channel attacks} are known to
leak sensitive information.

Recently, \CryptoBap{}~\cite{nasrabadi2023cryptobap,nasrabadi2025parallelfull} sought to bridge this gap by verifying cryptographic protocols at the machine-code level;
while binary-analysis platforms such as BINSEC~\cite{binsec} and BAP~\cite{bap} offer powerful program-analysis capabilities, they are
insufficient for such analyses, as they lack any mechanism
to semantically reconstruct protocol logic from binary.
Being in its early development stages, {\CryptoBap{}} suffers from limited scalability, restricting analysis to small or medium-sized protocols.
Its limitation mostly comes from the computational overhead of techniques like symbolic execution on large code bases and the difficulty of modeling dynamic protocol behaviors (e.g., session resumption, key rotation) in automated verifiers.
Moreover, \CryptoBap{} focuses mostly on trace properties in the Dolev--Yao model (e.g., secrecy, authentication) and does not account for hardware-induced side-channel vulnerabilities.
This oversight raises concerns, as side-channel leaks in protocol implementations can inadvertently expose secrets; however, so far side channels were not taken into account when analyzing protocol \emph{implementations}.
Traditional side-channel detection tools~\cite{scamv,guarnieri2021hardware,Osiris,cauligiConstantTimeFoundationsNew2020,danielHuntingHaunterEfficient2021}, which target cryptographic primitives or library code, fail to account for protocol-specific interactions that amplify leakage, such as timing variations that can happen during session establishment, error handling, order of operations, or memory-access patterns during key derivation.
Concretely, such tools can flag a secret-dependent branch, but not whether a protocol attacker can actually trigger it, nor whether the branched-on value carries a protocol-level secret in the first place, i.e., questions that require reasoning about the protocol, not only the code.

In this paper, we present a methodology for analyzing cryptographic protocol implementations, addressing both classical security properties and side-channel resilience.
Our approach extends \CryptoBap{}~\cite{nasrabadi2023cryptobap,nasrabadi2025parallelfull} with observation models~\cite{scamv,scamv:buiras2021} (a.k.a., leakage contracts~\cite{guarnieri2021hardware}) that enable automated analysis of real-world protocols against side-channel attacks.
Building on \CryptoBap{}'s core methodology---transpiling assembly code into an intermediate representation for symbolic execution and model extraction---we extract \Sapic models directly from protocol binaries that can be analyzed by \DeepSec for side channel leak and by \Tamarin/\ProVerif for reachability properties.

{\bf Threat model and scope.}
We consider an active network attacker in the Dolev--Yao model and a co-resident attacker (process) that can observe microarchitectural effects captured by our leakage contracts (e.g., instruction- or data-cache Prime+Probe measurements).
For speculative leakage, we adopt a Spectre-PHT-style model in which conditional branches may be transiently mispredicted and wrong-path memory accesses may become observable through side channels.
Our goal is \emph{not} to model all microarchitectural channels, but to lift well-scoped observation models to protocol verification and to expose security violations that arise from control flow divergence and memory-access behavior.

Our observation models can, in principle, capture ``classical'' constant-time violations such as secret-dependent table lookups (e.g., in an AES implementation) because secret-dependent load addresses become observable.
However, we abstract crypto library calls to focus on protocol logic and on protocol-level leakages.

As a case study, we conduct the first formal verification of the WhatsApp Desktop binary by extracting a model of its session management protocol (Sesame) and double ratchet mechanism.
WhatsApp is the most widely used messaging platform, with over 3 billion users across the globe~\cite{whatsapp}.
While Sesame has been formally analyzed at the specification level~\cite{cremers2023a}, no prior work extracted and verified the Sesame logic as \emph{implemented} by WhatsApp's client.
Given that WhatsApp is closed-source and similar systems have historically shown vulnerabilities (e.g., EternalBlue, \texttt{CVE-2017-0144}), this lack of implementation-level analysis is concerning.

For large applications like WhatsApp, considering all memory operations and function calls yields a large extracted model.
Even with simplification, state-of-the-art protocol verifiers still fail to terminate when analyzing models of this scale.
For reachability-style properties on WhatsApp (e.g., forward secrecy and post-compromise security), we therefore rely on \CryptoBap{}'s original extraction pipeline, which abstracts memory effects and cryptographic primitives in a way that is sound for finding trace-property violations under its assumptions.
For side-channel and privacy analyses, we apply our observation-aware extraction to smaller, protocol-relevant components where memory- and control-flow observations are essential and still scalable.

Using our models, we (i) prove forward secrecy of WhatsApp's session establishment and message exchange, (ii) confirm that a clone attacker can break post-compromise security---as previously identified for Signal application~\cite{cremers2023a}---and (iii) identify a novel privacy attack that leverages instruction-cache leakage during session establishment to infer whether two users have been in contact.
To summarize, our contributions are as follows:
\begin{enumerate}
    \item We extend \CryptoBap{} with leakage contracts and a model extraction path to \DeepSec to enable automated reasoning about protocol security in the presence of microarchitectural side channels.
    \item We provide the first formal model of WhatsApp's implementation of session management and double ratchet component extracted from its binary, and use \Tamarin/\ProVerif to prove forward secrecy and to confirm a post-compromise security break against a clone attacker.
    \item Applying our side-channel-aware analysis to WhatsApp session establishment, we find a privacy leak: a side-channel attacker can distinguish first-time from existing contacts, enabling them to infer the victim’s social graph. We also confirm the known unlinkability break in BAC.
\end{enumerate}

\noindent
\textbf{Responsible disclosure:}\quad The vulnerabilities found in WhatsApp Desktop were responsibly disclosed to Meta in March 2025. Meta confirmed that the application is vulnerable to identified attacks.

	\section{Background}

We build on \CryptoBap{}~\cite{nasrabadi2023cryptobap,nasrabadi2025parallelfull}, a binary analysis framework that extends security protocol verification to machine code to eliminate the need to trust compilers.
\CryptoBap{} extends the HolBA framework~\cite{DBLP:journals/scp/LindnerGM19} to verify ARMv8 and RISC-V machine code crypto protocols. It achieves this by extracting a formal model of the protocol under analysis, which can then be translated into models suitable for automated verification using \ProVerif, \CryptoVerif, \Tamarin, and \DeepSec.
In this section, we provide an overview of the \CryptoBap{} structure and introduce the necessary preliminaries to understand the contributions of this paper.

\subsection{HolBA Framework}
\begin{figure}
\adjustbox{varwidth=\linewidth,scale=0.9}{%
\begin{equation*} 
\begin{split}
	\birprog \in \fstlangcol{{prog}} & \bnfdef  \fstlangcol{{block}}^* \\
	\fstlangcol{{block}} & \bnfdef  ({\var{v},\fstlangcol{{stmt}}^*}) \\
	\var{v} \in \birval & \bnfdef  \mathit{string} \bnfsep \mathit{int}\\
	\fstlangcol{{stmt}}  & \bnfdef \fstlangcol{halt} \bnfsep \fstlangcol{jmp}(\var{e}) \bnfsep \fstlangcol{cjmp}(\var{e},\var{e},\var{e})\\  & \ \ \ \ \ \ \bnfsep \fstlangcol{assign}(\mathit{string}, \fstlangcol{e}) \bnfsep  \fstlangcol{assert}(\var{e})  
\\	\var{e} \in \birexp & \bnfdef  \var{v}\;  \bnfsep  \fstlangcol{\Diamond_u} \; \var{e}\; \bnfsep \var{e} \; \fstlangcol{\Diamond_b} \; \var{e} \bnfsep \fstlangcol{var} \ \mathit{string}\; \bnfsep \fstlangcol{load}(\var{e},\var{e},\! \mathit{int})\\ 
	 &
	\begin{array}{l}
		\;\;\;\; \bnfsep \fstlangcol{store}(\var{e},\!\var{e}, \var{e},\! \mathit{int}); \bnfsep  \fstlangcol{ifthenelse}(\var{e}, \var{e},\var{e})
	\end{array} 
\end{split}
\end{equation*}
}
\caption{$\birsymb$'s syntax}
\label{fig:birSyntax}
\end{figure}

HolBA~\cite{DBLP:journals/scp/LindnerGM19} is a library for binary analysis based on the HOL4 theorem prover~\cite{hol4} and the L3 specification language~\cite{foxl3modelsweb}. HolBA achieves this by transpiling binary code into the Binary Intermediate Representation ($\birsymb$
\footnote{We use \fstlangcol{co}\sndlangcol{lo}\com{rs} for different languages: \fstlangcol{RoyalBlue, \ math \ bold} for $ \birsymb $ and $ \sbirsymb $, and \sndlangcol{RedOrange, \ sans \ serif} for $ \Sapic $. Common elements use \textit{black italics}.}),
a simple, architecture-agnostic language designed to facilitate binary analysis and tool development.
To ensure soundness, the transpilation process is verified to preserve the semantics of the input machine code.

\cref{fig:birSyntax} depicts \birsymb{}'s syntax.
A $\birsymb$ program $\birprog$ consists of uniquely labeled blocks, with each block containing a sequence of statements.
Labels correspond to specific locations in the program and are commonly used as the target for jump instructions.
$\birsymb$ statements include
(a) $\fstlangcol{assign}$, to assign a $\birsymb$ expression to a variable,
(b) jumps (i.e., $\fstlangcol{jmp}$ or $\fstlangcol{cjmp}$),
(c) $\fstlangcol{halt}$, which serves as the termination instruction, and
(d) $\fstlangcol{assert}$, which evaluates a boolean expression and terminates execution if the assertion fails.
Expressions in $\birsymb$ include constants, variables, conditionals (i.e. $\fstlangcol{ifthenelse}$), arithmetic operations, denoted by $\fstlangcol{\Diamond_b}$ for binary and $\fstlangcol{\Diamond_u}$ for unary, as well as memory operations such as $\fstlangcol{load}$ and $\fstlangcol{store}$.

HolBA provides a proof-producing symbolic execution for $\birsymb$ \cite{lindner2026forward} which \CryptoBap{}~\cite{nasrabadi2023cryptobap,nasrabadi2025parallelfull} uses in its pipeline. This symbolic execution formalizes the symbolic generalization of $\birsymb$ (called $\sbirsymb$) to explore all execution paths of the program.
The symbolic semantics align with concrete semantics, enabling guided execution that maintains a set of reachable states arising from an initial symbolic state.
HolBA's symbolic execution allows for verifying functional correctness,
    but not (directly) protocol security, as it
lacks a suitable attacker model and concurrent behavior.
\CryptoBap{} bridges this gap by extracting formal models of the protocols from their implementations.
This model is then used to analyze security properties using external protocol verifiers.
{
\CryptoBap{} extracts formal models of protocols into two distinct modified versions of the applied $\pi$-calculus: one suitable for automated verification using \ProVerif and \CryptoVerif, proposed in~\cite{nasrabadi2023cryptobap}, and another for automated verification with \ProVerif, \Tamarin, and \DeepSec, utilizing the \Sapic toolchain introduced in~\cite{nasrabadi2025parallelfull}.
}
Since our goal is to check \textit{trace equivalence}~\cite{DBLP:journals/jcs/McLean92}, we have chosen to extract \Sapic models, as \Sapic backends support equivalence properties.

\subsection{\Sapic \& \DeepSec}
\begin{figure}[t]
	\adjustbox{varwidth=\linewidth,scale=0.9}{%
		\centering	
		\[
		\begin{array}{ll ll}
			\langle \sndlangcol{P},\sndlangcol{Q}\rangle~::= \\[2mm]
			~~~~~  \sndlangcol{0}  &
			~~|~~\sndlangcol{!P}\\
			~~|~~\pin(x);~\sndlangcol{P}  &
			~~|~~\sndlangcol{P \mid Q} \\
			~~|~~\pout(x);~\sndlangcol{P} &
			~~|~~\pnew~n;~\sndlangcol{P} \\ 
			~~|~~\pif~\sndlangcol{\phi}~\pthen~\sndlangcol{P}~\pelse~\sndlangcol{Q} &
			~~|~~\pevent~{e};~\sndlangcol{P} \\
			~~|~~ \plet~t_1 = t_2~\pin~\sndlangcol{P}~\pelse~\sndlangcol{Q} \\
		\end{array}
		\]
	}
	\caption{A fragment of the syntax of \Sapic process calculus.
		In this figure,
		$e, t_1,t_2 \in \Term$, 
		$n \in \privnames$, 
		$x \in \Vars$.	}
	\label{fig:SAPICsyntax}
\end{figure}

The Dolev--Yao model includes an attacker that exploits logical flaws in a protocol, but cannot compromise cryptographic primitives~\cite{DBLP:conf/focs/DolevY81}.
In this model, the cryptographic primitives are considered perfect---for instance, guessing a key is impossible.
A specific set of rules defines the abstract manipulation of messages, while other alterations are not permitted.

In a protocol execution, messages are represented as terms.
High-entropy values are modeled by constants derived from an infinite set of names $\allnames$ classified into public names $\pubnames$ (available to attackers) and secret names $\privnames$.
Terms denoted as $\Terms$  are constructed using names derived from $\allnames$, variables sourced from a variable set $\Vars$, and by applying function symbols from $\Funcs$ on terms.

\Sapic~\cite{cheval2022sapic+} is a dialect of applied $\pi$-calculus that provides a language that soundly translates to \Tamarin~\cite{meier2013tamarin}, \ProVerif~\cite{blanchet2001efficient} and \DeepSec \cite{cheval2018deepsec}.
\Sapic enhances \SapicOriginal~\cite{kremer2016automated} by introducing destructors and $\plet$ bindings with pattern matching and $\pelse$ branches.
A fragment of \Sapic's syntax is shown in~\cref{fig:SAPICsyntax}.
The $\pin$ and $\pout$ constructs are responsible for receiving and outputting messages through {a channel visible to the attacker.}
The $ \pevent $ construct is used to raise events that pertain to the reachability properties of the model.
Furthermore, the $\pnew$ construct enables the generation of new values.
Conditionals are described by first-order formulae $\sndlangcol\phi$ over equalities on terms, possibly containing variable quantifiers, as in~\cite{kremer2016automated}.
The $\Sapic$ syntax features \textit{stateful} processes that modify globally shared states, which are excluded in~\cref{fig:SAPICsyntax} for simplicity.

\Sapic facilitates the analysis of equivalence properties through its backends, specifically \DeepSec, a specialized tool designed for this purpose.
\DeepSec focuses on indistinguishability properties, particularly trace equivalence.
It employs a language similar to \ProVerif but without the ``$ \sndlangcol{!} $'' operator
for unbounded replication,
as it supports only bounded verification.
Unlike \ProVerif, \DeepSec provides a decision procedure that guarantees termination, given sufficient resources.
As a result, \DeepSec can effectively check trace equivalence in cases where \ProVerif fails to terminate, though it requires bounding the number of replications.

\paragraph{\bf Backend protocol verifiers.}
In our pipeline, \Sapic serves as a front-end that can be discharged to different backends depending on the property of interest and the required level of automation.
\ProVerif is typically effective for reachability properties (e.g., secrecy and authentication) and supports unbounded replication, but it may not terminate on complex models.
\Tamarin supports richer equational theories and an interactive proof mode to guide proof search with lemmas and proof hints.
\DeepSec provides a terminating decision procedure for bounded verification of process equivalences such as trace equivalence, which we use to encode privacy properties (e.g., unlinkability~\cite{arapinis2010analysing}) and conditional non-interference~\cite{guanciale2020inspectre} in the presence of side-channel observations.

\subsection{Side Channels \& Observational Models}
\label{sec:background:sc}
Resource sharing is inevitable in computing due to limitations in available resources. However, if not done carefully, it can introduce unintended information flow channels, also known as
side channels.
These channels can potentially be exploited by a malicious process to exfiltrate secret information from trusted ones.

Attacks that exploit the data and instruction caches are among the most commonly used side-channel attacks~\cite{Tsunoo03cryptanalysisof,Aciicmez:2006:TCA:2092880.2092891, Neve:2006:AAC:1756516.1756531, Tromer:2010:ECA:1713125.1713127}.
One widely used technique for extracting information
via caches is known as Prime+Probe~\cite{Percival2005}.
In an instruction-cache attack using this technique, first, the attacker \textbf{primes} the cache by filling it with their own instructions. Then, while the victim executes, some of the attacker's cached instructions may be evicted. Finally, the attacker \textbf{probes} the cache by \textit{measuring access times} to their instructions to detect evictions that reveal the victim's execution behavior.

The number of attack techniques exploiting microarchitectural features, like caches, to leak secret data continues unabated.
Consequently, the study of information flow analysis techniques to ensure the absence of information leakages due to side channels is a topic of increasing relevance.
A formal model of side channels is essential for such an analysis.
However, explicitly modeling all the intricate features of modern processors---like cache hierarchies, replacement policies, and memory interactions---is almost infeasible due to their complexity and because
many microarchitectural details are not publicly available.
To address this challenge, \textit{abstract observational models}~\cite{scamv,scamv:buiras2021}  (a.k.a., \emph{leakage contracts}~\cite{guarnieri2021hardware}) provide an alternative by overapproximating an attacker's capabilities.

An observation model
$\obsmodel$
extends the abstract representations of a processor's operational semantics
by a set of system states $\mstates$,
a set of possible attacker observations $\obs$ and a
labeled transition relation $\rightarrow_m \subseteq S \times \obs \times S$
indexed with the execution mode $m \in \{r, t\}$.
When $m=r$, we mean that the processor executes at the software-visible ISA level using a sequential transition system, while with
$m=t$, we denote the transition relation of some target microarchitecture where the information flow may be affected by optimizations such as out-of-order or speculative execution.
Essentially, observations define which parts of the processor state influence the side channel at each transition. This enables information flow analysis without requiring us to know the exact microarchitectural behavior.

The primary property to formalize the absence of {microarchitectural} leakages due to side channels is \textit{conditional non-interference}~\cite{guanciale2020inspectre}. Let $\mstate \in \mstates$ be a system state, including microarchitectural components like caches, $\functionsymbol{\traces}\!: \exectraces \mapsto 2^\obs$ be a function to extract the sequence of observations from a given execution trace $\exectrace{} \in \exectraces $, and $\mstate \sim_{\obsmodel} \mstate'$ is the state's indistinguishability relation w.r.t. the model $\obsmodel$. Then we say:

\begin{definition}[Conditional non-interference (CNI)]\label{def:conditional-non-interference}
    A system is \emph{conditionally non-interferent} if
    for all  indistinguishable initial states $\mstate$ and $\mstate'$
(i.e., $\mstate \sim_{\obsmodel} \mstate'$),
	if for every execution sequence $\exectrace{r}_1 =\mstate \xrightarrow{o_1}_{r} \mstate_1 \dots \xrightarrow{o_n}_{r} \mstate_n$ there exists a corresponding sequence
	$\exectrace{r}_2 = \mstate' \xrightarrow{o'_1}_{r} \mstate'_1 \dots
	\xrightarrow{o'_{n}}_{r} \mstate'_{n}$ such that $\traces(\exectrace{r}_1) = \traces(\exectrace{r}_2)$, then for every execution
	$\exectrace{t}_1 = \mstate \xrightarrow{o_1}_{t} \mstate_1 \dots
	\xrightarrow{o_n}_{t} \mstate_n$, there must also exist a corresponding
	$\exectrace{t}_2 = \mstate' \xrightarrow{o'_1}_{t} \mstate'_1 \dots
	\xrightarrow{o'_n}_{t} \mstate'_n$ such that
        $\traces(\exectrace{t}_1) = \traces(\exectrace{t}_2)$.
\end{definition}

A common strategy to prevent cache timing side channels in the literature is the \emph{constant time} (CT) policy~\cite{barthe2014system}, which requires that memory accesses and control flow decisions should depend only on public (non-secret) information.
In this paper, the observational model $\ctobsmodel$ formalizes this policy and it makes the program counter of each instruction and the accessed memory addresses observable.

Alas, speculative execution introduces new attack vectors that break the assumptions of CT execution.
\emph{Spectre} attacks~\cite{kocher2019daniel} exploit speculation to leak data through {side channels like caches}. These attacks are characterized by a speculative primitive that allows leaking secrets during speculative execution.
We use the observational model $\specobsmodel$ proposed by Buiras et al.~\cite{scamv:buiras2021} to capture \emph{Spectre-PHT/Spectre-V1-style} leakage, i.e., leakage caused by transient execution along a mispredicted conditional branch.
The model introduces \textit{refined} (a.k.a., \textit{shadow}) observations that represent operations executed transiently on the misspeculated path.

\begin{figure}
  \includegraphics[width=0.9\linewidth]{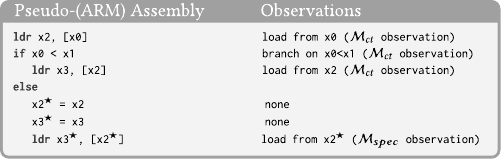}
  \caption{The Spectre V1 example instrumented via $\ctobsmodel$ and $\specobsmodel$. We marked shadow observations with $\star$.}
  \label{fig:shadowobs}
\end{figure}

{
\subsubsection*{Observation refinement}
Technically, an observation model $\obsmodel$ groups states into equivalence classes where states appear indistinguishable.
Observation refinement improves this partitioning by introducing a refined model $\obsmodel'$ that further partitions these classes. Essentially, $\obsmodel'$ captures additional behavioral variations, particularly those linked to side-channel effects, that $\obsmodel$ may overlook.
For instance,~\cref{fig:shadowobs} depicts the Spectre V1 primitive annotated with the attacker's observation from the $\ctobsmodel$ model and shadow observation from the $\specobsmodel$ that enable the attacker to observe an operation that may execute during the speculation.}
The current implementation of our toolchain focuses on branch misprediction; modeling other Spectre variants (e.g., store-to-load forwarding in Spectre-v4, or return-stack-buffer effects) would require developing suitable observation models.
More generally, our framework is parametric in the observation model: a new model requires only formalizing which processor-state components are observable for each instruction type and implementing the corresponding instrumentation, while the rest of the pipeline is unchanged.

	\begin{figure*}[t]
    \centering
	\includegraphics[width=0.9\linewidth]{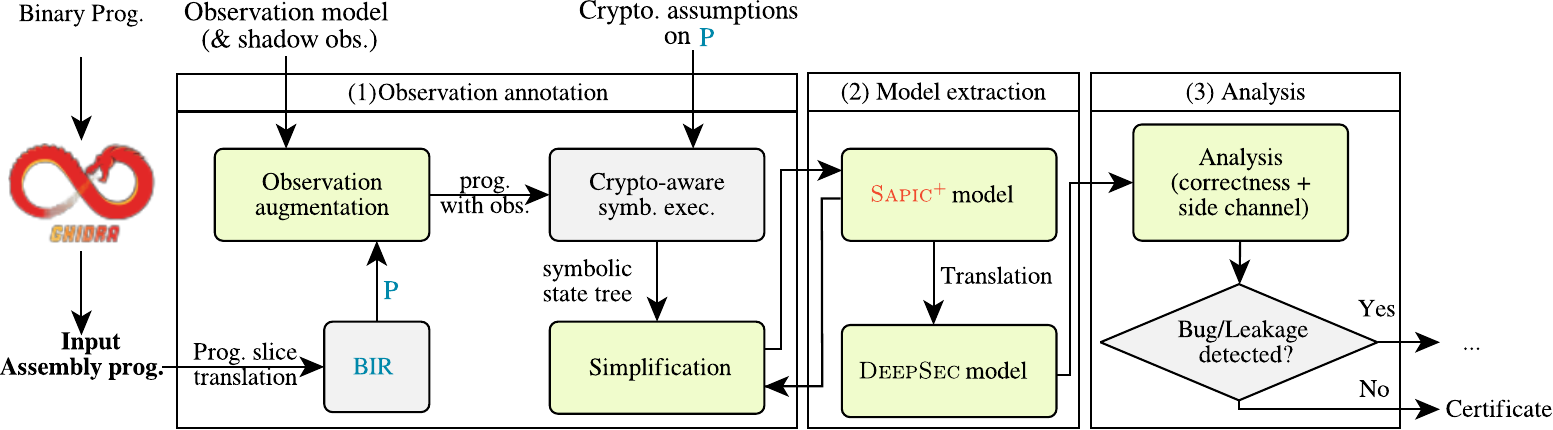}
        \caption{Organization of our approach; new features are in green. We reuse \CryptoBap{}'s model-extraction core, but add observation instrumentation, simplification rules, and a \DeepSec backend for leakage detection.
    }
	\label{fig:overview}
\end{figure*}

\section{Methodology}
\label{method}

To reason about crypto protocols' resilience to hardware side-channels, we combine symbolic execution, observational models, and protocol verification techniques---the source code of our framework is available at~\cite{source}.
As~\Cref{fig:overview} illustrates, our approach begins
{with reverse-engineering the executable binaries using Ghidra~\cite{ghidra}, after which we transpile the resulting assembly code}
into $\birsymb$ using the HolBA framework~\cite{DBLP:journals/scp/LindnerGM19}.
We then annotate the $\birsymb$ program with attacker observations and symbolically execute it to explore all execution paths. We simplify and translate the resulting symbolic execution tree into a formal model in \Sapic calculus, which we then further simplify to improve reasoning scalability. Finally, we translate this model into formats compatible with automatic analysis of side-channel resilience using \DeepSec.

In the following, we first describe how we annotate $\birsymb$ with observational models to capture side-channel leaks. Next, we detail our symbolic execution engine, tailored for dealing with crypto primitives and hardware interactions. We then explain the extraction and simplification of the protocol model. We conclude by discussing how we leverage \DeepSec to analyze the simplified \Sapic models for side-channel resilience.

\begin{figure}[t]
	\includegraphics[width=0.85\linewidth]{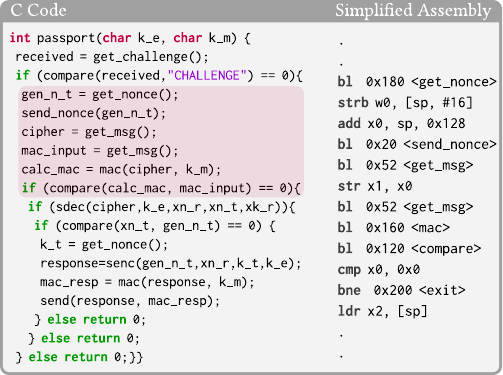}
	\caption{Running example. The assembly snippet corresponds to the highlighted C code.}
	\label{fig:run-exp-c}
\end{figure}

\begin{figure*}[t]
	\includegraphics[width=1\linewidth]{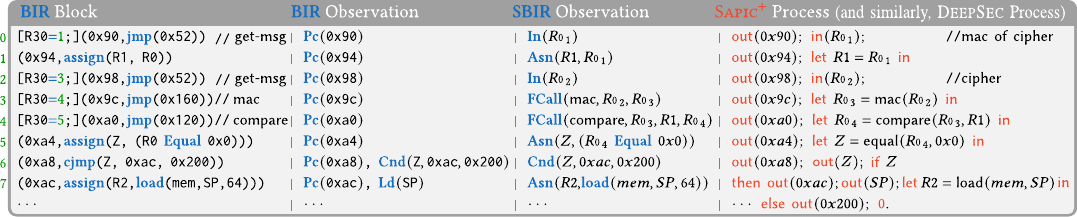}
	\caption{Excerpt of the $\birsymb$ blocks of the BAC protocol { in~\cref{fig:run-exp-c}}, together with the corresponding $\birsymb $ observations, symbolic execution events and extracted \Sapic model. $0x200$ is the address of the $\birsymb$ $\fstlangcol{halt}$ statement, which translates to $\sndlangcol{0}$. The \Sapic process then translates to the \DeepSec process, which closely mirrors what is illustrated here. Jumps (at lines 0, 2, 3, and 4) are the translation of \textit{branch and link} instruction used for function calls in ARM, which requires updating the \textit{link register} $ R30 $. We present this register update in $[..]$ to mean that it is not relevant to what we intend to present in this example.}
	\label{fig:run-exp-bir}
\end{figure*}

\paragraph{\bf Running example}
We use the Basic Access Control (BAC) protocol as a running example.
\Cref{fig:run-exp-c} shows a C implementation of the BAC's session-establishment logic---developed in-house since production e-passport code is not publicly available---along with the simplified assembly fragment used throughout this section.
Our implementation follows the BAC protocol and complies with the relevant International Civil Aviation Organization standards specifications~\cite{force2004pki}.
BAC is a three-pass challenge-response protocol for mutual authentication between an e-passport and a reader.
The reader first sends a challenge; the e-passport responds with a fresh nonce.
The reader then samples its own nonce, encrypts both nonces under the pre-shared encryption key $k_e$, and transmits the ciphertext together with a MAC computed under $k_m$.
Upon receipt, the e-passport verifies the MAC, decrypts the ciphertext, and checks that its nonce is present. If all checks succeed, the reader is authenticated (and the e-passport is authenticated to the reader analogously).
In \Cref{fig:run-exp-bir}, we highlight the corresponding $\birsymb$ blocks to illustrate the steps of our methodology.

\subsection{Reverse Engineering}

Analyzing real-world, closed-source binaries requires first isolating the protocol logic inside a larger code base. In WhatsApp Desktop, the code that implements the Signal/Sesame protocol stack constitutes only a small fraction of the overall executable, yet it is intertwined with application logic, storage, and networking.

\paragraph{\bf Binary extraction}
We use Ghidra~\cite{ghidra} to (i) locate protocol-relevant entry points, (ii) follow cross-references and call graphs to collect dependent functions, and (iii) export the \emph{exact instructions} of the selected region.
We start from functions that (a) retain symbols or strings associated with the Signal protocol stack and (b) interact with crypto-library entry points and network I/O, which can be recognized using approaches in the spirit of~\cite{meijer2021s}.
We then expand this set by following data and control dependencies (slicing) to obtain a region that covers message parsing/serialization, state updates, and calls to cryptographic primitives.

\paragraph{\bf Correctness considerations.}
We treat the reverse-engineering step as part of the trusted computing base: if the selected region omits relevant dependencies, the extracted model may be incomplete.
To mitigate this, whenever we cannot resolve a dependency (e.g., an external library call), we conservatively overapproximate it by modeling its outputs as fresh symbolic inputs from the environment.
Importantly, we do \emph{not} rely on Ghidra's decompiler for semantic reasoning, i.e., after exporting instructions, all semantic lifting and symbolic reasoning are performed on HolBA's verified $\birsymb$ semantics.
Finally, speculation is accounted for \emph{after} lifting by instrumenting $\birsymb$ with shadow observations. Thus, we do not require the decompiler to predict speculative paths.

\subsection{\birsymb{} with Observation Models}
To integrate side-channel leakage into our analysis, we instrument $\birsymb$ programs by annotating each $\birsymb$ block with attacker observations, which are expressions that describe information that may leak through side channels. These observations can be conditional, meaning they only occur if specific conditions hold in the current state. We adopt established observational models from the Scam-V platform~\cite{scamv}.
Our $\birsymb$ \emph{observations} $\birevent{} \in \birobs$ include:
\begin{center}
	$\birevent{} \vcentcolon\vcentcolon=	\oempty \ \mid \ \ld{\fstlangcol{a}} \ \mid \ \st{\fstlangcol{a}} \ \mid \ \cnd{\nodecond,\fstlangcol{a_1},\fstlangcol{a_2}} \ \mid \ \opc{\fstlangcol{a}}$
\end{center}

Observation $\opc{\fstlangcol{a}}$ exposes the label of the $\birsymb$ block, which corresponds to the program counter,
$\cnd{\nodecond,\fstlangcol{a_1},\fstlangcol{a_2}}$ reveals the outcome of the conditional branch's condition and exposes the addresses of each branch, and
$\ld{\fstlangcol{a}}$/$\st{\fstlangcol{a}}$ exposes the address operand of $\fstlangcol{load}$/$\fstlangcol{store}$ instructions.
All other instructions are considered non-leaking and emit the empty observation $\oempty$.

Depending on the observation models ($ \ctobsmodel$, $ \specobsmodel $, or any other), the observations assigned to each $ \birsymb $ block can be either normal or shadow observations (see~\cref{sec:background:sc}).
E.g., take the last line in~\cref{fig:shadowobs}, the $ \specobsmodel $ observation annotated to the $ \birsymb $ block of the corresponding assembly code is $ \ld{\mathsf{x2}^\star} $. Our observation instrumentation is not formally proved sound against a given observational model, but it closely follows (we reused partially) the Scam-V framework, whose validation against real hardware is established.

Compared to other platforms, Scam-V implements a more detailed observation of $\fstlangcol{load}$ and $\fstlangcol{store}$ statements.
While we simplify the presentation, we preserve the same level of granularity.

The $\birsymb$ transition relation $\birtrans{\birevent{}}{}{}{\!} \subseteq \birstates \times \birevents \times \birstates$ defines how states evolve while releasing observations. Here, $\birstates$ is a set of $\birsymb$ states, and $\birevents$ denotes the set of observations. Starting from an initial state $\birstate{0} \in \birstates$, a sequence of transitions produces a $\birsymb$ observation trace $\birtrace = \birevent{1} , \dots , \birevent{m}$.
\Cref{fig:run-exp-bir} illustrates a $\birsymb$ program snippet (first column) and its annotated observations (second column).

\subsection{Symbolic Execution}
To analyze side-channel leakages across all feasible paths, we extend \CryptoBap{'s} symbolic execution to track attacker observations alongside path constraints. Each symbolic state $\sbirstate{} \in \sbirstates$ now encapsulates a path condition---a logical constraint indicating a condition under which a path is taken---and a list of symbolic observations which represent leaked information (e.g., memory addresses and branch targets) that an attacker could infer.

Let $\sbirtrans{\sbirevent{}}{}{}{\!} \subseteq \sbirstates \times \sbirevents \times \sbirstates$ denote the transition relation of $\sbirsymb$ and let an $\sbirsymb$ trace be a sequence of observations such that $\sbirtrace = \sbirevent{1} , \dots , \sbirevent{m}$.
$\sbirsymb$ observations $\sbirevent{} \in \sbirevents$ include $\birsymb$ observations on symbolic values or expressions, as well as observations related to network communication and calls to crypto primitives, event functions and random number generation.
$\Input{x}$ represents the incoming message $x$ from the environment, whereas $\Output{x}$ denotes a message $x$ dispatched to the environment.
$\freshv{n}$ indicates when the program generates a random number $n$, and  $\fcall{\functionsymbol{f},x_1,\dots,x_m,y}$ refers to a function call $\functionsymbol{f}$ with inputs $x_1,\dots,x_m,$ and an output $y$.
In addition, $\event{e}$ signifies the occurrence of a visible event $e$, $\Loopsym$ denotes the start of a loop, and $\assign{x}{\var{e}}$ represents the assignment of the $\birsymb$ expression $\var{e}$ to the variable $x$.
The $\sbirsymb$ observations of our example are depicted in~\cref{fig:run-exp-bir} (third column).

\subsection{Model Extraction}\label{sec:model-extraction}

To analyze protocol implementations for side-channel leakages, we symbolically execute the instrumented program and derive a symbolic execution tree $\tree$ from the $\sbirsymb$ execution, with root denoting the initial symbolic state. We use $\tree$ to extract the protocol's \Sapic model. The tree captures all feasible execution paths explored during symbolic analysis, annotated with attacker observations and protocol-specific events.

The execution tree $\tree$ consists of
(a) leaves ($\leafnode$), indicating the end of a complete execution path in the tree, i.e., where the $\fstlangcol{halt}$ statement is encountered,
(b) event nodes ($\node(\nodeevent)\bnfconcat\tree'$) each containing a sub-tree $\tree'$ and an event $\nodeevent$,
and
(c) branch nodes ($\branchingnode(\cnd{\nodecond, \fstlangcol{a_1}, \fstlangcol{a_2}},\fstlangcol{T_1}, \fstlangcol{T_2})$) each with  the condition $\nodecond$ and subtrees $\tree_\fstlangcol{i}$ with their respective addresses $\fstlangcol{a_i}$. 

Our symbolic execution generates two successor states for the nodes representing a branching statement (i.e., $ \fstlangcol{cjmp}$), and we continue constructing subtrees from these states.
For other statements' nodes, we get an event node with either one or no successor in the tree. {The exceptions are the nodes containing \textit{indirect jumps} which may have multiple successors that are discovered iteratively using an SMT solver following the approach outlined in~\cite{nasrabadi2023cryptobap}.}

Having constructed $\tree$, we extract the protocol model by translating $\tree$ into its \Sapic model using the rules in~\cref{fig:sbirtosapic}.
The leaves in the tree are translated into a null process $\sndlangcol{0}$, and events $\nodeevent$ in the event nodes are translated into their corresponding \Sapic constructs.
{For example, we translate the attacker observations $ \ld{\fstlangcol{a}} $, $ \st{\fstlangcol{a}} $, and $ \opc{\fstlangcol{a}} $ within event nodes to the $ \pout(\sbirtoiml{\fstlangcol{a}}) $ constructs.
The observations $ \cnd{\nodecond, \fstlangcol{a_1}, \fstlangcol{a_2}} $ are preserved in the branching nodes of $\tree$, where each condition $\nodecond$ is a symbolic $\birsymb$ expression translated into an equivalent \Sapic term.
To maintain the attacker’s observations during simplification (cf.~\cref{sec:simp}), we first disclose the translation of $\nodecond$ to the attacker before translating the branching node into the \Sapic conditional construct, where $ \sbirtoiml{\nodecond} $ serves as its condition.
Each branch of this construct includes the translation of the corresponding addresses $\fstlangcol{a_i}$—also revealed to the attacker—and the translation of respective subtrees $\tree_\fstlangcol{i}$.
}
\Cref{fig:sbirtosapic} presents the standard rules for translating expressions.

{The more interesting case is the translation of function applications used, e.g., to translate memory load/store and bitwise operations.}
For instance, a $ \fstlangcol{load}(mem,a,l)$, where $l \in \{1, 8, 16, 32, 64, 128\}$, translates to $\functionsymbol{load}(mem,a)$, with $mem$ and $a$ representing symbolic values for the memory and the address respectively.

The fourth column in~\cref{fig:run-exp-bir} shows the extracted \Sapic process of our example.
An attacker capable of measuring microarchitectural states
can detect if the MAC check fails (by observing pc $0x200$) or succeeds
(by observing program counter $0xac$).
However, this is not visible without accounting for side-channel observations, as the program halts when the MAC check fails.
\begin{figure}
\adjustbox{varwidth=\linewidth,scale=0.9}{%
\begin{equation*} 
	\begin{aligned}[t]
\begin{split}
  \tree& = \!\leafnode \!\bnfsep \!\node(\nodeevent)\bnfconcat\tree'\!\bnfsep \branchingnode(\cnd{\nodecond, \fstlangcol{a_1}, \fstlangcol{a_2}},\fstlangcol{T_1}, \fstlangcol{T_2}) \ \text{event tree} \\
  & \sbirtoiml{\leafnode} \hspace{1.4cm} \mapsto\ \hspace{0.1cm} \sndlangcol{0}\\
  & \sbirtoiml{\node(\nodeevent)\ \bnfconcat\tree'} \bnfdef \hspace{2.5cm} \text{events nodes} \\
  &
 \begin{array}{lll}
 \sbirtoiml{\node(\oempty)\bnfconcat\tree'} & \mapsto & \sbirtoiml{\tree'}\\
 \sbirtoiml{\node(\sbirevent{})\bnfconcat\tree'}, \sbirevent{}\in \{ \ld{\fstlangcol{a}},\st{\fstlangcol{a}},\opc{\fstlangcol{a}} \}\!\!\!\!
& \mapsto & \pout(\sbirtoiml{\fstlangcol{a}});\sbirtoiml{\tree'} \\
  \sbirtoiml{\node(\event{{e}})\bnfconcat\tree'} & \mapsto & \pevent \ {e};\sbirtoiml{\tree'}\\
  \sbirtoiml{\node(\Input{x})\bnfconcat\tree'} & \mapsto & \pin (x);\sbirtoiml{\tree'}\\
  \sbirtoiml{\node(\Output{x})\!\bnfconcat\!\tree'}\! & \mapsto & \pout(x);\sbirtoiml{\tree'}\\
  \sbirtoiml{\node( \assign{x}{\fstlangcol{e}})\bnfconcat\tree'} & \mapsto & \plet \ x = \sbirtoiml{\fstlangcol{e}} \ \pin \ \sbirtoiml{\tree'}  \\
  \sbirtoiml{\node(\freshv{n})\bnfconcat\tree'} & \mapsto & \pnew \ {n};\ \sbirtoiml{\tree'}\\
  \sbirtoiml{\node(\Loopsym)\bnfconcat\tree'} & \mapsto & \sndlangcol{!} \ \sbirtoiml{\tree'}\\
  \sbirtoiml{\node(\fcall{\functionsymbol{f},x_1,\ldots,x_n,y})\bnfconcat\tree'} & \mapsto & \plet \ y=\functionsymbol{f}(x_1,\ldots,x_n) \\ && \pin \ \sbirtoiml{\tree'} \ \pelse\ \sndlangcol{0} \\
\end{array}\\
  & \sbirtoiml{\branchingnode(\cnd{\nodecond, \fstlangcol{a_1}, \fstlangcol{a_2}},\fstlangcol{T_1}, \fstlangcol{T_2})}  \mapsto \\ & \hspace{2cm} \pout(\sbirtoiml{\fstlangcol{\nodecond}});\pif \ \sbirtoiml{\fstlangcol\phi}\ \pthen \ \pout(\sbirtoiml{\fstlangcol{a_1}});\sbirtoiml{\fstlangcol{T_1}} \ \pelse \ \pout(\sbirtoiml{\fstlangcol{a_2}});\sbirtoiml{\fstlangcol{T_2}}\\
  & \sbirtoiml{\fstlangcol\phi \in \birexp} \bnfdef \hspace{3.2cm} \text{Expressions} \\
  &
 \begin{array}{lll}
  \sbirtoiml{\var{b} \in \birval}  & \mapsto & \constrans{b}\in \pubnames \\
  \sbirtoiml{\fstlangcol{var} \ x }  & \mapsto & {x}\in \Vars\\
  \sbirtoiml{\fstlangcol{\phi_1 \Diamond_b \phi_2 }}  & \mapsto & \sbirtoiml{\fstlangcol{\Diamond_b}}(\sbirtoiml{\fstlangcol{\phi_1}},\sbirtoiml{\fstlangcol{\phi_2}}) \hspace{0.55cm} \text{Binary operations}\\
  \sbirtoiml{\fstlangcol{\Diamond_b}} &\mapsto & \left\{\begin{array}{ll}                                     
                                          \functionsymbol{equal}   & \fstlangcol{Equal} \\
                                          \functionsymbol{plus}, \functionsymbol{mult}, \ \dots & \fstlangcol{Plus}, \fstlangcol{Mult}, \ \dots
                                        \end{array}
                                 \right.\\
  \sbirtoiml{\fstlangcol{\Diamond_u \phi'}}          & \mapsto & \sbirtoiml{\fstlangcol{\Diamond_u}}(\sbirtoiml{\fstlangcol{\phi'}}) \hspace{1.45cm} \text{Unary operations}\\
  \sbirtoiml{\fstlangcol{\Diamond_u}} &\mapsto & \left\{\begin{array}{ll}
                                        \functionsymbol{not}            & \fstlangcol{Not} \\ 
                                        \bot            & \text{otherwise}
                                        \end{array}
                                 \right.     \\       
 \sbirtoiml{\fstlangcol{f}(\fstlangcol{e_1},\dots,\fstlangcol{e_m})}  & \mapsto & \sbirtoiml{\fstlangcol{f}}(\sbirtoiml{\fstlangcol{e_1}},\dots, \sbirtoiml{\fstlangcol{e_m}})                  
\end{array}    \\   
\end{split}
\end{aligned}
\end{equation*}
}
\caption{Translating $\tree$ into a $\Sapic$ model:  $ {e}, {x}, x_1,\ldots,x_n,y \in \Vars $ are variables, $ n \in \privnames $ is a secret name, and $ \functionsymbol{not}$,  $\functionsymbol{equal}$,  $\functionsymbol{plus}$,  $\functionsymbol{mult}$,  $\functionsymbol{f} \in \Funcs$ are function symbols.  
$ \sbirevent{} $ and conditional observations are either normal or shadow observations.
} 
\label{fig:sbirtosapic}
\end{figure}

\subsection{Simplification}
\label{sec:simp}

The extracted models include memory operations and attacker-observable events, and are therefore too large for current protocol provers.
To reduce model complexity, we used abstractions in $\birsymb$, pruned paths at $\sbirsymb$, and applied several simplification rules at the \Sapic level, which are shown in~\cref{simp-rules}.

At the $\birsymb$ level, we introduce a \emph{storage abstraction} akin to
KLEE's treatment of files, pipes, and terminals~\cite{cadar2008klee}.
We abstract the SQLite engine by modeling the effects of database creation, the table schema, and the read and write operations on the database file.
In particular, WhatsApp stores session keys in the SQLite database.

HolBA’s semantic-preserving transpiler inserts assertions into $\birsymb$ programs that encode \emph{well-formedness invariants} of executions, e.g., after each stack operation it adds $\fstlangcol{assert}(sp_{\mathit{low}}\!\le\! sp \!\le\! sp_{\mathit{high}})$ to confine the stack pointer to the current frame. By construction, each inserted $\fstlangcol{assert}(\chi)$ is an invariant for executions of the original binary that start from well-formed initial states. Consequently, any $\sbirsymb$ path that violates such an assertion is infeasible.
We therefore cause the path to fail at that point and prune its suffix. This does not affect our side-channel analysis because the inserted assertions hold for all executions from well-formed states, and our observation semantics are guarded. Hence, assertion-violating paths cannot produce observations.
We call such assertion-violating paths \emph{infeasible}, and the ``infeasible'' column of tables~\ref{fig:obs-cases} and~\ref{fig:wa-cases} counts them. These differ from the \emph{unreachable} branches handled next.

At the $\sbirsymb$ level, further simplifications are possible. An example is pruning paths that the SMT solver marks as unreachable, like branches of conditionals that are always false. While for functional correctness analysis, eliminating such paths is possible, we cannot apply this simplification when we look for side-channel leaks, as unreachable branches can leak during speculative execution.

At the \Sapic level, nil processes indicate the end of a complete path.
When a let-binding `$\plet~x = e~\pin~\sndlangcol{P}~\pelse~\sndlangcol{Q}$' results in nil processes in both its success and failure branch (i.e., $\sndlangcol{P}=\sndlangcol{0}$ and $\sndlangcol{Q}=\sndlangcol{0}$), it can be removed and substituted with $\sndlangcol{0}$.
Moreover, for a conditional that involves paths following the same sequence of constructs, like $ \pif~\sndlangcol{\phi}~\pthen~\sndlangcol{P}~\pelse~\sndlangcol{P}$, the conditional can be replaced with the corresponding constructs in those paths, $\sndlangcol{P}$.
In order to ensure that this simplification would not affect our side channel analysis, we preserve the observations related to the conditional (if any) and add them to the simplification result.

Live variable analysis is a data-flow analysis used in compiler optimization to identify live variables at every point in a program~\cite{alfred2007compilers}.
A variable $x$ is live at a given point $p$ in the program if $x$ will be used along some path starting from $p$.
This analysis becomes especially important when we want to eliminate a $\plet$ construct.
We use this technique to determine whether the variable $x$ used in the $\plet~x = e~\pin~\sndlangcol{P}$ construct remains alive within the sub-process $ \sndlangcol{P}$.
If so, we retain the $\plet$ construct, otherwise, we substitute it to $ \sndlangcol{P}$.

Finally, observational models like $\ctobsmodel$  require making the program counter of instructions (i.e., $ \opc{\fstlangcol{a}} $) visible to the attacker. When extracting a formal model of a program annotated with such observational models, we generate an
$\pout(\sbirtoiml{\fstlangcol{a}})$ for each instruction (see~\cref{fig:sbirtosapic}). However, the large number of output constructs makes \DeepSec computation expensive without a significant benefit, mainly when they originate from the same branch.
Therefore, leaking the program counters of conditional branches is sufficient and other program counter leakages can be eliminated.
This greatly simplifies the extracted \Sapic process (e.g., see~\cref{fig:obs-cases}).
\begin{table}[]
\adjustbox{varwidth=\linewidth,scale=1}{
	\begin{tabular}{l|l|l|l}
		\Sapic  & \pif~\sndlangcol{\phi}~\pthen~\sndlangcol{P} & \plet~\sndlangcol{\phi}~\pin~\sndlangcol{0} & 	\plet~$ x = e $~\pin~\sndlangcol{P}~s.t. \\
		Process & \hspace{2em}\pelse~\sndlangcol{P}  & \hspace{2em}\pelse~\sndlangcol{0} &	$x \notin $  \vars(\sndlangcol{P})  \\     \hline
		Simplified &  \sndlangcol{P}  &  \sndlangcol{0} &  \sndlangcol{P}
	\end{tabular}
}
\caption{Simplification rules: $\vars$ are a set of variables for a given process.}
\label{simp-rules}
\end{table}

\subsection{Detecting {Leakages} with \DeepSec}

We use the protocol verifier \DeepSec~\cite{cheval2018deepsec} to detect side-channel leakage exploitable by a Dolev--Yao network attacker who can additionally observe the leakage events produced by our contracts.
\DeepSec specializes in privacy properties expressed as trace equivalence: given two processes $P_1$ and $P_2$, it checks whether an attacker can distinguish their observable execution traces.

For instance, in BAC, an attacker can exploit observable control-flow differences (e.g., $\pout(0xac)$ versus $\pout(0x200)$ in~\cref{fig:run-exp-bir}) to infer information about message structure.
By replaying previously observed messages, the attacker can distinguish successful from unsuccessful MAC verification and thereby violate BAC's intended privacy guarantees.

\DeepSec is well-suited for such analyses because it supports equivalence checking.
Formally, it decides whether $P_1 \approx P_2$ holds for the class of processes it supports; when it terminates, it either proves equivalence or returns a counterexample in the form of distinguishing traces.

To isolate leakage introduced by speculative execution, we evaluate conditional non-interference (\cref{def:conditional-non-interference}) by running the same equivalence check under two observation models.
Concretely, we check equivalence once under the constant-time observations $\ctobsmodel$ and once under the refined speculative observations $\specobsmodel$.
If $P_1 \sim_{\ctobsmodel} P_2$ but $P_1 \not\sim_{\specobsmodel} P_2$, the additional distinguishing power comes from speculation, witnessing a speculative-only leak.

A counterexample under $\ctobsmodel$ (resp.\ $\specobsmodel$) yields two symbolic traces distinguishable by the chosen observation model; under the assumptions of our lifting and abstractions, this corresponds to a concrete binary-level side-channel distinguisher within our leakage contract.
Conversely, when \DeepSec reports equivalence, the result is bounded (due to bounded replication) and depends on the soundness of our abstractions for cryptographic primitives, external I/O, and the selected observations.

	\section{Evaluation}
\label{sec:case-study}

In our evaluation, we separate two goals that require different backends and abstractions:
(i) \emph{reachability-style} security analysis of WhatsApp Desktop (e.g., forward secrecy under a Dolev--Yao network attacker), and
(ii) \emph{equivalence-style} privacy and side-channel analysis (unlinkability and conditional non-interference) under the observation models $\ctobsmodel$ and $\specobsmodel$.
For (i), we reuse \CryptoBap{}'s established extraction pipeline and discharge the resulting models to \ProVerif and \Tamarin.
For (ii), we apply our observation-aware extraction (\cref{method}) and analyze the resulting \Sapic processes with \DeepSec.

\Cref{fig:wa-cases} summarizes the sizes and extraction/verification costs for the WhatsApp components (Sesame session handling and double ratchet).
\Cref{fig:obs-cases} summarizes the corresponding statistics for the side-channel case studies (BAC and WhatsApp session establishment) under $\ctobsmodel$ and $\specobsmodel$.
The reduction percentages report the relative decrease from the raw extracted model (\textit{all}) to the simplified model (\textit{simp}). Note that verification time may increase because simplification and backend translation add overhead.

\paragraph{\bf Manual vs.\ automatic effort.}
Once a protocol-relevant region is selected, the pipeline (lifting, observation instrumentation, symbolic execution, extraction, simplification, translation, and verification) runs automatically.
The manual inputs are: (a) selecting entry points and defining the analysis boundary in Ghidra, (b) deciding which external calls are treated as cryptographic abstractions or environment stubs, and (c) stating the verification query for the chosen backend.
For BAC, this required a few hours of reverse engineering and modeling.
For WhatsApp, locating and validating protocol-relevant entry points in Ghidra required a few days; after that, the remainder of the pipeline was fully automated.%

\subsection{Evaluation of WhatsApp  with \CryptoBap{}}

WhatsApp allows users to exchange messages, share status posts, and make audio and video calls.
Since 2016~\cite{whatsapp2024whitepaper}
WhatsApp uses a modified version of the Open Whisper Systems' Signal protocol as the basis for end-to-end encryption.
This encryption protocol prevents WhatsApp's servers and other third parties from accessing the plaintext of user messages or calls.
Messages are encrypted with ephemeral cryptographic keys that are
regularly updated (using ratcheting or a new handshake), preventing
 attackers from decrypting previously transmitted messages, even if the current encryption keys are compromised.

We extract a formal model of WhatsApp's binary to verify forward secrecy and post-compromise security.
For this part, we use \CryptoBap{}'s original extraction pipeline to obtain a \Sapic model that can be discharged to \ProVerif{} and \Tamarin{} for reachability-style analysis; this analysis is orthogonal to the side-channel observations studied later in this section.
We reverse-engineered the WhatsApp Desktop's binary using Ghidra.
Unlike the iOS or Android versions,
function symbols were not stripped from the
desktop application,
allowing us to match the function names with the latest version of the \texttt{libsignal}, Signal's protocol implementation.

\subsubsection{Components}
\label{sec:component}

The double ratchet protocol enables secure (confidential and authentic)
message exchange between the two parties.
It builds on
the Extended Triple Diffie-Hellman (X3DH) key agreement protocol~\cite{signal2025specifications},
which allows parties to establish a shared secret key
using mutual authentication based on their public keys.

Initially, the
key is obtained via X3DH, and called the \emph{root key}.
Subsequent ephemeral keys are generated
using ratcheting, {so called} because
earlier ephemerals cannot be derived from later ones.
{Ratcheting is} \emph{asymmetric} if the parties switch roles (i.e., from sender to receiver or vice versa) and \emph{symmetric} if they maintain the same role.
Symmetric ratcheting provides \emph{forward secrecy}{:} even if long-term secrets (the secret keys to the public keys used in the handshake) or future ephemerals are revealed, past ephemerals and thus the messages sent with them remain secret.
Asymmetric ratcheting establishes fresh ephemerals that are unknown to the
attacker even if \emph{past} ephemerals are known, unless the attacker
actively attacks the handshake.
This provides \emph{post-compromise} security: even after ephemeral keys are
revealed, the authenticity and confidentiality of future ephemerals
(and the messages encrypted with them) can be recovered.

 Sesame defines the session management and operates on the
layer above the double ratchet protocol. It manages multiple double
ratchet sessions between the different devices associated with each user account. E.g., if Alex has $n_A$ devices and Blake has $n_B$ devices,
on each of Alex's devices, Sesame manages $n_B$ connections to Blake and
$n_A$-1 connections to Alex's other devices.
Sesame manages the creation, deletion, and use of sessions, maintaining a local database that records each party's devices and their associated sessions.
To establish a pairwise encrypted conversation between two users,
Sesame manages an instance of the double ratchet protocol (including X3DH key agreement)
between each of their devices.

\subsubsection{Extracted Model}\label{subsec:ext-mdl}
\begin{table*}
	\centering
	\resizebox{2.1\columnwidth}{!}{
		\begin{tabular}{ll|c|c|c|c|c|c|c|c|ccccc}
			\multicolumn{2}{c|}{\multirow{2}{*}{WhatsApp Protocol}} 
			& \multirow{2}{*}{\begin{tabular}[c]{@{}c@{}}\#ARM \\instructions\end{tabular}} 
			& \multicolumn{2}{c|}{\#symb.\ exec.\ paths} 
			& \multicolumn{3}{c|}{{\#$\Sapic$ \hspace{2.5em} LoC}} 
			& \multirow{2}{*}{\begin{tabular}[c]{@{}c@{}}\#TM\\LoC\end{tabular}} & \multirow{2}{*}{\begin{tabular}[c]{@{}c@{}}\#PV\\LoC\end{tabular}} 
			&  \multicolumn{3}{c|}{{time \hspace{1em} to \hspace{1em} extract}} 
			& \multicolumn{2}{c}{Overall verif. time} \\
			\multicolumn{2}{c|}{} &  & feasible & infeasible  & all & simplified & crypto. &  & & \multicolumn{1}{c|}{all} &  \multicolumn{1}{c|}{simplified} &  \multicolumn{1}{c|}{crypto.} & \multicolumn{1}{c|}{TM} &  PV \\ \hline
			\multicolumn{2}{l|}{(1) Initiate Session} & 6844 & 12 & 25 & 106 & 61 (42\% $\downarrow$) & 33 & \multirow{4}{*}{161} & \multirow{4}{*}{171} & \multicolumn{1}{c|}{13} &  \multicolumn{1}{c|}{14 (7\% $\uparrow$)} & \multicolumn{1}{c|}{12} & \multicolumn{1}{c|}{\multirow{4}{*}{19.76}} & \multirow{4}{*}{0.057} \\
			\multicolumn{2}{l|}{(2) Respond to (1)} & 5803 & 106 & 413 & 718 & 92 (87\% $\downarrow$) & 12 &  &  & \multicolumn{1}{c|}{65} &  \multicolumn{1}{c|}{60 (7\% $\downarrow$)} & \multicolumn{1}{c|}{38} & \multicolumn{1}{c|}{} &\\
			\multicolumn{2}{l|}{(3) Send Message} & 3041 & 156 & 368 & 983 & 429 (56\% $\downarrow$) & 166 &  &  & \multicolumn{1}{c|}{71} &  \multicolumn{1}{c|}{100 (40\% $\uparrow$)} & \multicolumn{1}{c|}{65} & \multicolumn{1}{c|}{} & \\
			\multicolumn{2}{l|}{(4) Receive Message} & 4181 & 7293 & 19322 & 31257 & 2033 (93\% $\downarrow$) & 360 & & \multicolumn{1}{c|}{} & \multicolumn{1}{c|}{26903} &  \multicolumn{1}{c|}{40128(49\% $\uparrow$)} & \multicolumn{1}{c|}{25782} & \multicolumn{1}{c|}{} & 
		\end{tabular}%
	}
	\caption{WhatsApp protocol analysis. The table includes, the size of the code under analysis, the explored symbolic paths, the extracted models---complete and simplified, consisting solely of cryptographic operations, and translated into \Tamarin and \ProVerif---along with the duration (in seconds) of each step. TM and PV refer to \Tamarin and \ProVerif, respectively.}
	\label{fig:wa-cases}
\end{table*}

We abstract the
X3DH key agreement protocol
with a private channel that `magically' communicates a master
secret key (initial root key) between two parties.
Like in the handwritten \Tamarin model for Signal
in~\cite{cremers2023a}, this avoids verification issues
with \Tamarin's limited Diffie-Hellman (DH) theory (see
\cite[Sec.  2.3 and 2.4]{cheval2022sapic+}
for further discussion on the support of DH theories in \Tamarin and
other protocol verifiers).

Excluding X3DH, we extract a model of the core components:
\begin{enumerate}
    \item \label{itm:component1} \emph{initiating a new session} in which the client requests the recipient’s keys from the WhatsApp server and calculates a session key,
    \item \emph{responding to a request to initiate a new session} by the recipient and calculating the corresponding session key,
    \item \emph{transmitting messages} in a session (symmetric ratchet), and
    \item \emph{receiving messages} within a session (asymmetric ratchet).
\end{enumerate}

\Cref{fig:wa-cases} presents the data obtained from our evaluation of the WhatsApp components using \Tamarin and \ProVerif.

By considering all memory operations and function calls (but abstracting cryptographic library calls), we initially obtain a \Sapic model with nearly 33{,}000 LoC. Simplification (\cref{sec:simp}) reduces this to about 2{,}600 LoC, but both \Tamarin and \ProVerif still fail to terminate at this scale.
For the reachability analysis reported below, we therefore use \CryptoBap{}’s memory-abstraction variant (i.e., without explicit load/store primitives). This variant yields a substantially smaller \Sapic model that remains suitable for analyzing security properties under the Dolev–Yao attacker model (the corresponding model’s lines of code are presented in the ``crypto.'' column of~\Cref{fig:wa-cases}).
We highlighted the memory-aware numbers above to emphasize the scalability bottleneck of state-of-the-art protocol verifiers that currently prevents full memory-precise verification of large real-world binaries.

\subsubsection{Implementation vs. Specification}
We find that, in certain scenarios, the protocol implementation behaves differently from its specification in the WhatsApp security
white-paper~\cite{whatsapp2024whitepaper}, in particular the component responsible for decrypting incoming messages (i.e., the component number 4 in~\Cref{fig:wa-cases}).
The white paper specifies that the chain and root keys are updated upon receiving a response, yet it lacks a sufficient explanation of the ratcheting mechanism~\cite[p.15]{whatsapp2024whitepaper} (in contrast to the Signal documentation~\cite{signal2025specifications}).
According to WhatsApp, the ratcheting process involves each party calculating its next chain and root keys using a shared ephemeral secret, derived from the sender’s and receiver’s ephemeral keys, and a root key whose origin is not clearly defined.
Instead, our model indicates three distinct behaviors for ratcheting. The next chain key is either derived from: 
(a) the current chain key,
(b) the new chain key, calculated from fresh root and ephemeral keys, and
(c) the fresh chain key, calculated from the current root key and a new ephemeral key.

This highlights the strength of model extraction. Most protocol
models have to rely on the specification to be correct, even if the source is available, a thorough comparison is tedious and requires tool support (such as model extraction for the source language). Often, behaviors are underspecified (as in this case), and the modeler fills the gap. Sometimes source-code inspection helps fill the gap, but WhatsApp's source code is closed.
Ultimately, formal analysis results rely on the adequacy of the
model. Gaps through underspecification or misspecification can lead to
overlooking attacks.

\subsubsection{Authenticity with \Tamarin \& \ProVerif}
\label{sec:authenticity}
We used our extracted model to analyze WhatsApp's session management and double ratchet protocol with \Tamarin and \ProVerif.

Utilizing these tools, we verify the security properties of our extracted model by defining queries and checking the reachability of all events in the model.
Forward secrecy ensures that past session keys remain secure even if long-term keys are compromised.
Our analysis shows that \Tamarin and \ProVerif successfully verify \textit{forward secrecy} for two parties initiating a session and exchanging secret messages.
On the other hand, post-compromise security ensures that if an attacker gains access to a party’s state, the protocol can ``heal'' and prevent the attacker from compromising future sessions.
However, \textit{WhatsApp violates post-compromise security despite employing the double ratchet protocol when faced with a clone attacker.}
Our analysis revealed this major issue, which prior work~\cite{cremers2023a} had identified in the Signal application and \emph{suspected} in WhatsApp.
Cremers et al.~\cite{cremers2023a} have proposed secure mechanisms that offer stronger guarantees.
However, our analysis showed that WhatsApp does not implement any of the proposals, allowing a clone attacker to break post-compromise security.

Note, while the analysis of Signal's implementation benefits from direct access to the source code (the model in~\cite{cremers2023a} was manually crafted but informed by the code), our models were derived entirely from the binary.
Moreover, while \ProVerif automatically identifies the post-compromise security violation for WhatsApp, we employed heuristics to guide \Tamarin{}’s proof search and manually selected 303 proof steps from the available options.

\subsection{Side-Channel Analysis}

The second part of our evaluation demonstrates how side-channel analysis of protocol implementations can reveal hardware-induced leakages.
Although abstract formal models and implementations of cryptographic protocols are often formally verified against security properties, they have not been analyzed in the presence of hardware-induced leakages.
Our methodology addresses this gap by combining model extraction with leakage contracts, enabling automated analysis of side-channel leaks.
We focus on BAC and WhatsApp due to their widespread use in real-world scenarios and their critical security implications to evaluate their resilience against potential attacks.

In these case studies, we assume the attacker can influence protocol inputs, e.g., by replaying or crafting messages.
Additionally, the attacker can observe microarchitectural side effects captured by our observation models, including program counter and load/store addresses, as would be possible for a co-resident process performing cache attacks.
Thus, the detected distinguishers reveal violations that are \emph{triggerable} by network interaction but \emph{observable} through side channels on the victim device.

\subsubsection{Basic Access Control Protocol}
\begin{table*}
	\centering
		\begin{tabular}{ll|c|c|c|c|c|c|ccc}
			\multicolumn{2}{c|}{\multirow{2}{*}{Protocols}} 
                        & \multirow{2}{*}{\begin{tabular}[c]{@{}c@{}}\#ARM \\instructions\end{tabular}}
                        & \multicolumn{2}{c|}{\# symb.\ exec.\ paths} 
                        & \multicolumn{2}{c|}{\# $\Sapic$ LoC}
                      & \multirow{2}{*}{\begin{tabular}[c]{@{}c@{}}\# \DeepSec \\ LoC\end{tabular}} & 
                      \multicolumn{2}{c|}{time to extract} &
			 Verif. time in \\
			\multicolumn{2}{c|}{} &  & feasible & infeasible  & all & simplified &   & \multicolumn{1}{c|}{all} &  \multicolumn{1}{c|}{simplified} & \multicolumn{1}{c}{\DeepSec}  \\ \hline
			\multicolumn{2}{l|}{BAC with $\ctobsmodel$} & 176 & 5 & 69 & 242 & 83 (65\% $\downarrow$) & 205 & \multicolumn{1}{c|}{9} &  \multicolumn{1}{c|}{7 (22\% $\downarrow$)} & \multicolumn{1}{c}{1}  \\
			\multicolumn{2}{l|}{BAC with $\specobsmodel$} & 194 & 15 & 170 & 630 & 240 (61\% $\downarrow$) & 589 & \multicolumn{1}{c|}{22} &  \multicolumn{1}{c|}{18 (18\% $\downarrow$)} & \multicolumn{1}{c}{7}  \\
			\multicolumn{2}{l|}{Initiate Session with $\ctobsmodel$} & 6844 & 12 & 65 & 283 & 90 (68\% $\downarrow$) & 220 & \multicolumn{1}{c|}{43} &  \multicolumn{1}{c|}{44 (2\% $\uparrow$)} & \multicolumn{1}{c}{2}  \\
			\multicolumn{2}{l|}{Initiate Session with $\specobsmodel$} & 6899 & 120 & 679 & 2872 & 903 (68\% $\downarrow$) & 2191 & \multicolumn{1}{c|}{148} &  \multicolumn{1}{c|}{130 (12\% $\downarrow$)} & \multicolumn{1}{c}{5} 
		\end{tabular}%
	\caption{Case studies with observation models. Columns show: ARM code size, number of symbolic paths analyzed, extracted model size (raw, simplified, and after translation to \DeepSec), and time (seconds) per step.}
	\label{fig:obs-cases}
\end{table*}

Many countries are adopting electronic biometric passports, a.k.a. e-passports.
These passports encode the holder's digital information within a Radio Frequency Identification (RFID) chip for interaction with passport readers.
However, e-passports face several threats related to security and privacy, including skimming, cloning, and eavesdropping~\cite{hancke2011practical,avoine2016survey}.
To address these concerns, the  International Civil Aviation Organization (ICAO)
standardized
security mechanisms, among those, the
BAC protocol.
BAC's goal is to
ensure that only authorized parties can access personal information stored in passports' RFID.
A party (the reader) is authorized if they know a machine-readable code printed on
each e-passport, which the passport holder typically provides by
physically handing the passport to the reader or scanning it
themselves (if the reader is a device).

Despite the ICAO standard's goal to create an unlinkable BAC protocol,
some implementations of the protocol
permit reidentification of a passport holder.
Arapinis et al.~\cite{arapinis2010analysing} found that these implementations
output {different error messages if a replayed message is invalid due to an incorrect MAC or an incorrect nonce.}
As the second case can only occur when the message is replayed to the
same passport (because each interaction uses a fresh nonce), while the first can only occur if the message is replayed to a different
passport (as the MAC key is fixed per passport), this allows for
identification of the passport holder.
Subsequently, the error messages were unified, but Chothia and Smirnov~\cite{chothia2010traceability} find that all e-passports are still vulnerable to a variant of the attack where instead of the error message, the attacker measures the computation time of the passport’s response to distinguish the incorrect-MAC branch and the incorrect-nonce branch.

To show how our methodology finds such attacks on
critical systems, we implemented the BAC protocol ourselves.
While the machine code is not easily extractable from RFID chips for researchers (and the
legal implications are unclear), standardization bodies could request
access and run similar analyses.
As depicted in~\cref{fig:run-exp-c}, our implementation does not allow distinction by error message, but rather by side-channel leakage.
Using our methodology, we found that an attacker can use this leak
to distinguish between both types of failures and thus break
unlinkability.
The implementation is vulnerable under
both
$\ctobsmodel$ and $\specobsmodel$, see \Cref{fig:obs-cases}.

\subsubsection{Privacy Properties with \DeepSec}
\label{sec:privacy-deepsec}
To analyze the WhatsApp application's vulnerability to hardware-induced leaks, we begin with the $\birsymb$ translation of the session-initiation component defined in \cref{subsec:ext-mdl}, \cref{itm:component1}.
We annotate this $\birsymb$ program using both the constant-time model ($\ctobsmodel$) and the speculation model ($\specobsmodel$).
For each annotated $\birsymb$ program, we perform symbolic execution, extract the corresponding \Sapic model, and translate it into \DeepSec.
The evaluation results for the session-initiation component with \DeepSec are summarized in~\cref{fig:obs-cases}.

\DeepSec checks trace equivalence, which is typically used to formalize privacy properties such as unlinkability and voting privacy~\cite{cheval2018deepsec}.
As a baseline, we analyze the standard \textit{unlinkability} property for WhatsApp session establishment.
Informally, unlinkability requires that an attacker cannot distinguish between two executions in which Alice establishes a session with different communication partners (e.g., Bob versus Charlie).
In our bounded analysis, this standard unlinkability property holds for both extracted models, under $\ctobsmodel$ and $\specobsmodel$.

We then consider a refined, application-level privacy property that captures whether a session establishment corresponds to a first contact with a recipient or to a subsequent session within an already existing conversation.
Intuitively, an attacker should not be able to distinguish these two cases.
When analyzing this refined property, \DeepSec finds a distinguisher under $\ctobsmodel$ (and hence also under $\specobsmodel$).
The interpretation of this distinguisher and the resulting privacy attack are discussed in~\cref{subsec:privacy-attack}.

As in the rest of our analysis, we assume an honest WhatsApp server.
Communication between WhatsApp clients and the server---which is, in practice, protected by the Noise protocol framework~\cite[p. 35]{whatsapp2024whitepaper}---is modeled using private channels.

\subsubsection{Discovery of Privacy Attack}\label{subsec:privacy-attack}
The distinguisher reported by \DeepSec corresponds to a privacy attack that reveals whether a victim (i.e., a WhatsApp account holder) is establishing a secure session with a recipient for the first time or is establishing a new session for an existing conversation.

For every secure session establishment, the initiator retrieves the recipient's \emph{pre-key bundle} from the WhatsApp server.
This bundle contains the recipient's public identity key, public signed pre-key, and, when available, a public one-time pre-key (OTPK)~\cite{whatsapp2024whitepaper}. According to the WhatsApp security whitepaper~\cite{whatsapp2024whitepaper}, the recipient's public OTPK is used only once and is removed from the server after it has been requested to initiate the first-time conversation.
Consequently, the presence or absence of an OTPK in the retrieved pre-key bundle reflects whether the session establishment corresponds to a first contact or a subsequent session.
This observation leaks information about the victim's social interactions.
In particular, it reveals if the victim is contacting a recipient for the first time or is communicating with a recipient with whom a secure relationship has already been established.
The attack applies to both one-to-one conversations and group chats.
An attacker can repeatedly add selected users to groups that include the victim and use the side channel to infer whether prior communication relationships exist (see~\cref{subsec:attack}).
Repeating this process enables reconstruction of the victim's social graph, revealing sensitive information about the victim's relationships, interests, and activities.

Our analysis identified this attack automatically.
After annotating the $\birsymb$ program of the session-initiation component with the observational models, extracting the respective models, and translating them into \DeepSec, the tool detected a distinguisher associated with the conditional processing of the recipient's OTPK.
This control-flow dependency is observable through instruction-cache side channels and therefore violates the refined privacy property introduced in~\cref{sec:privacy-deepsec}.

\subsubsection{Attack Vector (Instruction Cache)}\label{subsec:attack}
In our model, the above attack requires the side channel to reveal the
program's control flow to learn the presence or absence of the
recipient's public one-time pre-key (OTPK).
Both observation models represent realistic attack vectors~\cite{scamv:buiras2021}; we will now outline how our attack
can be realized as an \emph{instruction cache attack}.
Such attacks exploit the timing differences in the processor instruction cache behavior to infer which program paths are executed without requiring direct access to the program counter or memory contents.

WhatsApp Desktop's assembly code includes a conditional branch that checks for the existence of the one-time pre-key in the pre-key bundle during session establishment. When the one-time pre-key is present (indicating a first-time conversation), a specific set of instructions is executed to process it, whereas a different program path is taken if the key is absent (indicating an existing conversation). These two distinct execution paths result in different patterns of instruction cache usage.
{In our proof-of-concept attack implementation, we used the Prime+Probe technique~\cite{Percival2005} to monitor the instruction cache by co-locating a process on the same CPU core as the WhatsApp Desktop application. Interested readers can find the details of our PoC in Appendix~\ref{subsec:poc}}.

In a Prime+Probe attack, the attacker first `primes' the instruction cache by executing their own code to fill specific cache sets.
Next, the attacker triggers the victim to establish a new encrypted session, e.g., by initiating a group chat with the victim and inviting the third person.
Then, the attacker `probes' the cache by measuring the time it takes to re-execute their code.
If the victim's execution of the pre-key processing code evicts the attacker's instructions from the cache (due to cache set conflicts), the probe will take longer, revealing that the `first-time conversation' path was taken. Conversely, if the cache remains largely untouched, it indicates the `existing conversation' path was executed. By mapping the WhatsApp application's instruction addresses to cache sets, the attacker can reliably distinguish between these two cases.
The instruction cache attack eliminates the need to observe the program counter in the clear or have privileged access to the victim's system memory.

\subsubsection*{Real-World Impact}
The vulnerabilities we discovered have a real-life impact.
Imagine that $V$ is a journalist connected with numerous sources,
including whistle-blower $T$. The attacker $R$ is a regime that
enforces its citizens to install a certain application which allows it
to mount instruction cache attacks.\footnote{This is plausible, e.g.,
	Russia~\cite{Russians}, Kazakhstan~\cite{egovernment}, Saudi Arabia~\cite{Absher} and Singapore~\cite{singpass} provide certain
	services like visa or passport requests, tax declaration, or social
	services exclusively
	via government-provided Smartphone Apps.
	In other countries, like Austria, Estonia, or Germany
	digitalized services with a strong App focus increase the pressure
	to install such Apps, often in connection to digital ID~\cite{eu-government}.
	Some states in the U.S.\ also increase the pressure to
	install Apps for social services~\cite{279902}.
	Many more countries had indirect enforcement of the use of Apps
	during the COVID-19 pandemic, e.g., to access public buildings or for
	citizens who tested positive~\cite{tracetogether}.}
The regime infiltrates some of $V$'s WhatsApp groups unrelated to their journalistic work.  $R$ convinces a moderator in this
group to include $T$ in a chat discussion. Using our attack, $R$
confirms that $V$ was in contact with $T$.

	\section{Related Work}
\label{sec:relatedwork}

\subsection{Protocol Verification}
The usage of formal methods to analyze protocols and verify properties like secrecy and authentication dates back to Lowe's work~\cite{loweAttackNeedhamSchroederPublickey1995}. Traditionally, domain experts translate protocol specifications into formal models to analyze high-level behavior using theorem proving~\cite{paulsonInductiveApproachVerifying1998,basin2021abstract}, model checking~\cite{loweAttackNeedhamSchroederPublickey1995,basinModelCheckingSecurity2018}, and symbolic analysis~\cite{blanchetSecurityProtocolVerifier2022, cheval2022sapic+, meierTAMARINProverSymbolic2013}. Tools like \ProVerif{}~\cite{blanchetSecurityProtocolVerifier2022} and \Tamarin{}~\cite{meierTAMARINProverSymbolic2013} are effective in verifying security properties. But, these specification-based models abstract away implementation details, potentially overlooking vulnerabilities introduced during execution, like hardware-induced leakages~\cite{lipp2020meltdown,kocher2020spectre} or compiler bugs~\cite{sidhpurwala2019security,simon2018you}.

To address discrepancies between specifications and implementations, techniques like fuzz testing~\cite{bohme2016coverage,ammann2024dy}, differential testing~\cite{mckeeman1998differential}, symbolic execution~\cite{aizatulin2015verifying}, code generation~\cite{cade2012computationally,acaySecureSynthesisDistributed2024}, deductive verification~\cite{cade2012computationally,lattuadaVerusPracticalFoundation2024,arquintSoundVerificationSecurity2022}, and type checking~\cite{bhargavan2010modular,DBLP:journals/ieeesp/BhargavanFK16} have been employed, which target the implementation of protocols. Model extraction techniques further aim to verify implementations in high-level languages like $F\#$~\cite{bhargavan2006verified,backes2010computationally}, Java~\cite{o2008using,jurjens2009automated}, C~\cite{goubault2005cryptographic,aizatulin2015verifying}, and Rust~\cite{lattuadaVerusPracticalFoundation2024}. Yet, the complexity of these languages limits their practicality.

\CryptoBap{} addresses these gaps by extracting formal models directly from protocol machine code, avoiding reliance on specifications or high-level code.
Our work builds on \CryptoBap{}, but extends the analysis along three axes: we (i) begin from the binary of real (closed-source) applications---not the one we compile ourselves, (ii) instrument the lifted binary with leakage contracts (e.g., constant-time and speculative observation models) and extract observation-aware models, and (iii) connect the extracted models to equivalence-based verification in order to reason about protocol properties (e.g., unlinkability) in the presence of microarchitectural observations.

This enables us to use the extracted models to assess (a) whether protocol properties are preserved in the presence of hardware leaks and (b) the existence of microarchitectural leakages through the protocol verifier \DeepSec.
Unlike \CryptoBap{}, we extract the memory operation models into \Sapic and apply simplification rules to scale the extracted models for analysis with protocol verifiers.

\subsection{Side-Channel Analysis of Crypto Libraries}

In recent years, side-channel detection, mitigation, and formal analysis have become active areas of research, with numerous papers published on the topic. Here, we do not intend to survey all works in this domain (an incomplete list can be found~\cite{DBLP:journals/csur/LouZJZ21,DBLP:conf/sp/CauligiDMBS22}); rather, we focus on a few selected works that are closer to our approach.

Crypto libraries have been the prime target of side-channel attacks, e.g., timing attacks. The constant-time (CT) paradigm mitigates these by ensuring control flow and memory accesses are independent of secret data under sequential execution~\cite{DBLP:conf/pkc/Bernstein06}. High-assurance frameworks like Jasmin~\cite{DBLP:journals/iacr/OlmosBCGLOSYZ24} and FaCT~\cite{DBLP:conf/sp/ShivakumarBBCCGOSSY23} use formal methods, such as information-flow type systems, to enforce CT and produce verified, efficient implementations.

However, Spectre attacks~\cite{kocher2019daniel}, exploiting speculative execution, challenge the CT paradigm by leaking secrets even in CT-compliant code. Research has extended CT to speculative constant-time (SCT) to protect against Spectre variants. Shivakumar et al.~\cite{DBLP:conf/sp/ShivakumarBGLOPST23} introduced a type system in Jasmin for SCT against Spectre-v1 with minimal overhead using selective speculative load hardening. They have also addressed declassification issues under speculation, proposing relative non-interference (RNI) and efficient countermeasures. Arranz Olmos et al.~\cite{DBLP:journals/iacr/OlmosBCGLOSYZ24} extended this to all Spectre variants, including Spectre-RSB, with low overheads. Other approaches like
Swivel~\cite{DBLP:conf/uss/NarayanDMCJGVSS21} and Serberus~\cite{DBLP:conf/sp/MosierNMT24} offer protections but may rely on hardware or incur higher overheads.

SCT-style work primarily targets CT guarantees for \emph{crypto primitive implementations} and often reasons about a concrete source language or assembly semantics.
Our work is complementary. We take observation models inspired by Scam-V/SCT-style semantics and lift them to \Sapic protocol models in order to analyze \emph{protocol-level} properties that can be violated by application (e.g., secret-dependent error handling or first-contact branches), even when cryptographic primitives are treated as ideal.
Throughout, we use ``hardware-induced'' leakage in the microarchitectural sense (e.g., cache-based observations), and do not model physical channels such as electromagnetic (EM) or power.

However, previous studies did not connect microarchitectural side-channel reasoning to protocol-level verification of real protocol \emph{implementations}
 (as opposed to specifications or isolated primitives); this is the gap we target.
In this work, we propose a methodology designed to identify such leakages and address this gap.

	\section{Discussions}
\label{sec:discussions}

\paragraph{\bf Soundness of proposed methodology}
Regarding our methodology soundness, transpilation from assembly to $ \birsymb $ and symbolic execution remain sound, although we lack formal proofs validating our translation to \Sapic (see~\cref{sec:model-extraction}).
Since our primary objective with the proposed methodology is attack detection, this is acceptable, as attacks can be manually validated if false positives remain manageable.
However, a positive security result from \DeepSec cannot be fully trusted.
For instance, we obtain a verification result for the strong secrecy of the master key (initial root key) derived during WhatsApp's session initialization with \DeepSec, but refrain from reporting it as we cannot be sure it would hold.

Symbolic execution is usually only shown sound, but it is possible
to show its completeness w.r.t. assignments consistent
with a path condition. If the subsequent translation step can be shown
to maintain that pattern, we should be able to show that a symbolic equivalence
between the translated processes implies a concrete trace equivalence
as long as  (symbolic) traces with consistent assignments are only
mapped to (symbolic) traces with consistent assignments. In our concrete application, this could be achieved by revealing the path constraints in the trace, as the program is compared to itself and thus the conditionals are observed through the program-counter observations.

\paragraph{\bf Component selection via Ghidra}
We currently identify WhatsApp components manually in Ghidra using (i) preserved function symbols corresponding to the Signal protocol, and (ii) interactions with the crypto library and network I/O, which can be systematically recognized~\cite{meijer2021s}.
In principle, information-flow analysis could automate this step by combining existing detection techniques~\cite{meijer2021s} with conservative over-approximation (treating all incoming inputs as arbitrary).
We are working toward integrating such automation.

\paragraph{\bf Mitigation against side-channel attacks}
Side-channel risks can be reduced either by constant-time rewrites (eliminating secret-dependent branches and memory access patterns) or by removing the cache channel (e.g., isolation/partitioning or cache flushes).
Implementations with effective isolation, partitioning, or flushing are not vulnerable under our observation-based model.
Our toolchain can validate \emph{code-level} mitigations by re-extracting the binary and re-running \DeepSec equivalence under $\ctobsmodel$ and then $\specobsmodel$.
However, we cannot currently verify fence-based mitigations or hardware/OS defenses; doing so would require extending the underlying semantics and observation models.

\paragraph{\bf Limitations of protocol verifiers}
The extracted models can reach the practical limits of current protocol verifiers.
Even hand-written and optimized models (e.g., for TLS~1.3~\cite{cremersComprehensiveSymbolicAnalysis2017}, WPA2~\cite{DBLP:conf/uss/CremersKM20}, and SPDM~\cite{DBLP:journals/iacr/CremersDN24}) may require hours to days of verification time, substantial memory, and extensive proof guidance.
These tools have improved on \emph{protocol complexity}, but not primarily on scaling with the \emph{size} of automatically generated rule sets.
Consequently, our methodology would benefit from better support for decomposition and modular verification, which remains limited in existing tools.

\paragraph{\bf Dependence on crypto libraries}
Our analysis is conducted in the Dolev--Yao model and therefore assumes crypto libraries are trusted.
If primitives deviate due to bugs, unexpected weaknesses, or their own side channels, our conclusions may not hold.

\paragraph{\bf Applicability constraints}
Our approach assumes access to the target binaries and reliance on a known cryptographic library.
In some settings (e.g., e-passport implementations), external researchers may not have access to code at all.
While WhatsApp employs basic obfuscation, we were still able to extract meaningful models; more targeted anti-analysis obfuscation could impede extraction, though such measures may themselves be suspicious.

\paragraph{\bf ISA and scalability limitations.}
Our implementation targets AArch64 and RISC-V because HolBA provides lifting for these ISAs.
Supporting additional ISAs (e.g., x86-64) requires a lifter with comparable semantic assurance.
Scalability is limited by symbolic execution and by the size of extracted \Sapic models, especially when modeling memory explicitly.
This motivates using a memory abstraction for full-program reachability in WhatsApp, and focusing observation-aware side-channel analyses on smaller components.

\paragraph{\bf Scope of side-channel modeling.}
Our leakage contracts expose a small observation set (program counter information, symbolic load/store addresses, plus speculative ``shadow'' observations).
This captures many cache-based attacks, but not all physical or microarchitectural effects. Thus, a found distinguisher indicates a concrete leak, whereas a proved equivalence does not exclude channels outside the observation set.

	\section{Concluding Remarks}
\label{sec:conclusion}

We presented a methodology for analyzing crypto protocol \emph{implementations} that combines binary-level model extraction with protocol verification under explicit leakage contracts.
Our approach extends \CryptoBap{} with attacker observation instrumentation, a Ghidra-assisted front-end for locating protocol code in real binaries, and an equivalence-checking backend (\DeepSec) for privacy and conditional non-interference.

We extracted an implementation-level model of WhatsApp Desktop's session management and double ratchet components and verified forward secrecy, while confirming a post-compromise security break against a clone attacker.
For side-channel, we found a new instruction-cache leak in WhatsApp session establishment that enables social-graph inference and confirmed the known unlinkability break in BAC under microarchitectural observations.

Our results suggest that protocol implementations deserve the same rigor of side-channel analysis traditionally applied to cryptographic libraries: protocol verifiers provide principled, composition-aware notions of secrecy and privacy, and lifting these notions to implementation traces makes it possible to detect protocol-level violations that are invisible in specification-only models.

\section*{Acknowledgments}
This work was partially supported by the Wallenberg AI, Autonomous Systems and Software Program (WASP) funded by the Knut and Alice Wallenberg Foundation. The first author was a PhD candidate at Saarland University during the preparation of this work. We also gratefully acknowledge gifts from Intel and Amazon.

    \bibliographystyle{ACM-Reference-Format}
    \balance
	\bibliography{refs,zotero}
	\appendix
	\section{Appendix}
	\subsection*{A. Proof-of-Concept Implementation}
\label{subsec:poc}
\begin{figure}[t]
	\centering
	\includegraphics[width=\columnwidth]{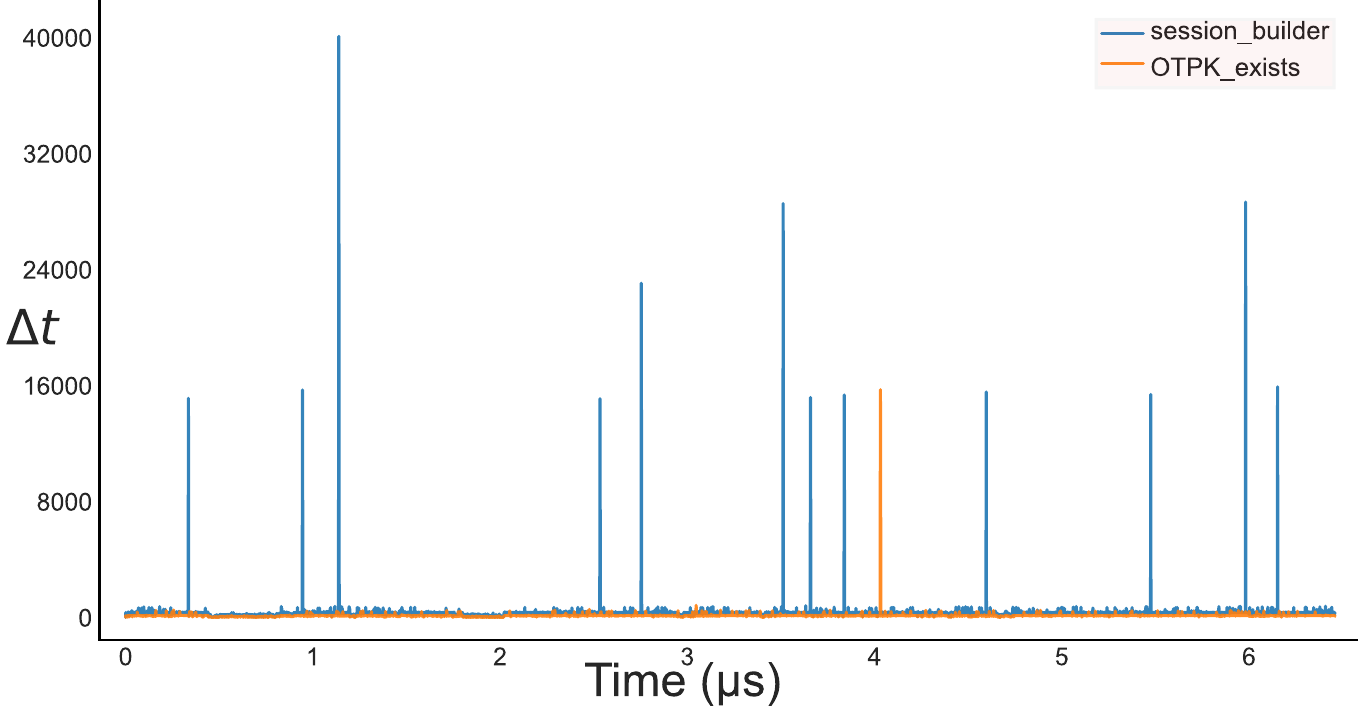}
	\caption{Observed access latency ($\Delta t$). \textit{session\_builder} is the address of the function initiating a secure session, and \textit{OTPK\_exists} the address of an instruction run when a one-time pre-key creates the session's master secret.}
	\label{fig:poc}
\end{figure}

To evaluate the feasibility of our attack in~\cref{subsec:attack}, we implemented a proof-of-concept targeting WhatsApp Desktop. It is based on the Flush+Flush framework~\cite{Flush+Flush}, modified for macOS, and is available at
\href{https://github.com/FMSecure/CryptoBAP/tree/ccs2026/HolBA/src/tools/parallelcomposition/examples}{\small Proof-of-Concept Implementation}. Experiments ran on a 2019 MacBook Pro with an Intel Core i7-9750H (6 cores, 2.6 GHz) and 16 GB RAM, running macOS Sonoma 14.1.2.

The attack leverages a Prime+Probe instruction cache side channel to infer whether the victim initiates a new conversation, i.e., whether a secure session is being established. The adversary controls a co-resident user-space process scheduled on the same physical core as the target. We assume the attacker knows the virtual addresses of two functions in the WhatsApp binary: one initiating a secure session, and another triggered when a one-time pre-key is present (\textit{session\_builder} and \textit{OTPK\_exists} in~\cref{fig:poc}, respectively).

In each cycle, the attacker primes the instruction cache by accessing other code mapping to these cache sets and lines. After a short delay allowing for potential victim execution, the attacker re-accesses the same addresses and measures the latency ($\Delta t$). Elevated latency indicates eviction, implying the victim executed instructions mapping to the same cache line; low latency suggests the lines remained untouched.

Our macOS-compatible iteration of Flush+Flush integrates timing-based cache probing via \textit{mach\_absolute\_time()}, which yields a high-resolution timestamp in platform-dependent units, enabling precise measurement of access latency. This straightforward access-timing technique infers potential cache evictions, improving the accuracy of cache-related timing.

\Cref{fig:poc} shows a time series of the latency $\Delta t$: the x-axis is time (in microseconds), the y-axis is the latency. Distinct spikes correspond to the victim executing one or both target code paths, revealing first-time conversation establishment (highlighted in orange).

\subsection*{B. Ethical Considerations}
Our work targets defensive security: extracting protocol models from executables and analyzing them under explicit microarchitectural leakage contracts.
\paragraph{Human subjects and data.}
Our WhatsApp analysis reveals a side channel inferring the victim's social graph, sensitive data whose unauthorized inference breaches privacy. We conducted no human-subject research and collected no personal data; all experiments ran offline on software and hardware under our control.
\paragraph{Responsible disclosure.}
We disclosed our findings to Meta with reproduction instructions. Meta confirmed the vulnerabilities, but we have no confirmation of a deployed fix.
\paragraph{Authorization.}
We analyze only lawfully obtained software and access no systems without authorization. Our reverse engineering of WhatsApp Desktop may conflict with Meta's license, but we disclosed responsibly upon discovery.
\paragraph{Societal impact.}
WhatsApp serves billions of users. By examining its binary, we close the gap between advertised guarantees and real-world protections, exposing privacy risks such as social graph inference that especially threaten journalists, activists, and users in repressive environments. Publishing such analyses under responsible disclosure pressures vendors to fix subtle flaws and raises accountability standards for deployed cryptographic systems.

\subsection*{C. Open Science}
Our framework's source code is available at~\cite{source}. Artifacts needed to evaluate our core contributions are linked below.
\begin{itemize}
\item \CryptoBap{}: \href{https://github.com/FMSecure/CryptoBAP/tree/ccs2026}{https://github.com/FMSecure/CryptoBAP}
\item \Tamarin (incl.\ \Sapic): \url{https://tamarin-prover.com/}
\item \ProVerif: \url{https://bblanche.gitlabpages.inria.fr/proverif/}
\item \DeepSec: \url{https://deepsec-prover.github.io/}
\item Ghidra: \url{https://github.com/NationalSecurityAgency/ghidra}
\end{itemize}

\subsection*{D. Use of Generative AI and LLMs}
This paper was refined for grammar, spelling, and minor style with Grammarly.
    
\end{document}